\documentclass[12pt]{article}
\usepackage{amssymb}
\usepackage{amsfonts}

\input{tcilatex}

\begin{document}

\begin{center}
\textbf{Quadratic Solitons in Negative Refractive Index Medium}

\bigskip

Andrew I. Maimistov$^{a}$ \footnote{%
electronic address: amaimistov@hotmail.com}

Ildar R. Gabitov$^{b}$, Elena V. Kazantseva$^{c}\bigskip $

$^{a}$ Department of Solid State Physics, Moscow Engineering Physics
Institute, Kashirskoe sh. 31, Moscow, 115409 Russia

$^{b}$Department of Mathematics, University of Arizona, 617 N. Santa Rita
Ave., Tucson, AZ 85721, USA

$^{c}$Laboratoire de Physique de l'Universit\`{e} de Bourgogne (L.P.U.B) Unit%
\`{e} Mixte de Recherche du CNRS 5027, Facult\'{e} des Sciences Mirande 9
Avenue Alain Savary BP 47 870, 21 078 Dijon Cedex, France

\bigskip

\textbf{ABSTRACT}
\end{center}

\bigskip

We are considered the propagation of the fundamental and second harmonic
solitary waves under the slowly varying envelope pulses approximation in a
quadratic nonlinear medium that characterized by negative refraction index
at the frequency of fundamental wave and by positive refractive index at the
second harmonic frequency. We find the some solutions of the evolution
equations which described the steady state simulton, coupled second harmonic
and fundamental frequency waves propagation in this materials.

\textit{PACS}: 42.65. Tg

\textit{Keywords}: Negative refraction, left-handed materials, parametric
interaction, optical harmonics

\bigskip

\section{Introduction}

In the past few years, new developments in structured electromagnetic
materials have given rise to negative refractive index materials which have
both negative dielectric permittivity and negative magnetic permeability in
some frequency ranges. These materials are often referred to as \emph{%
left-handed materials} (LHM) or materials with negative refraction (NRIM).
The properties of such materials were analyzed theoretically by Mandelschtam
and Veselago many years ago \cite{M1,M2}. It is very important that the wave
vector is anti-parallel with the Poynting vector inside LHM \cite{M1}-\cite%
{M7}. The occurrences of the LHM in microwave region have been demonstrated
experimentally in the \cite{M8}-\cite{M11}. The experimental evidence for
the existence of NRIM in optical region was found \cite{M12}-\cite{M16}
also. The results of investigation and discussion of the applications of
negative refractive medium represent in reviews \cite{M23,M24}.

Unusual features of the NRIM manifest itself in the passing the interface
between this material and the positive refractive index one. Also, the
refractive index of the same medium can be positive in one frequency region
and to be negative in another frequency region. Hence the features of NRIM
would be made itself evident in interaction of wave packets having carrier
frequencies from different spectral regions, where refractive indexes have
opposite signs \cite{R25, R26}. Second harmonic generation (SHG) is first
example where negative refraction leads to distinctly different pictures of
spatial distributions of the interacting waves intensity \cite{R5, R6}. Here
the quadratic nonlinear NRIM acts similar to distributed Bragg reflector,
i.e., Bragg gratings. Namely, the harmonic wave propagates towards
fundamental (pump) wave. In nonlinear Bragg grating under some conditions
the connected pair of the incident and reflected waves develops. It is gap
soliton propagating in nonlinear grating \cite{R30}. We can expect that in
quadratic nonlinear NRIM the second harmonic and fundamental wave pulses are
able to produce the complex steady state solitary wave that is similar to
quadratic soliton \cite{R3, R4} or simulton \cite{R33}.

In this paper the propagation of the complex steady state of the fundamental
wave and second harmonic wave considered. It is assumed that refractive
index for the frequency of fundamental wave is negative, and refractive
index at the harmonic wave frequency is positive. The particular solutions
of the equations describing the evolution of these waves are founded. They
are analogues of the solitons \cite{R3, R4} and cnoidal waves \cite{R34, R35}
in quadratic nonlinear medium. Under condition that value of the phase
mismatch is very large the propagation interacting waves approximately
governs by nonlinear Schr\"{o}dinger equation \cite{R36, R37}.

\bigskip

\section{Constituent equations}

Consider the propagation of the pulses of parametrically coupled waves,
i.e., fundamental wave (or pump wave) and second harmonic one. The type I
phase-matching condition is taking place \cite{R1, R2}. In this case the
slowly varying envelopes of pump and harmonic waves evolve according to
following system of equations can be written in following form

\begin{eqnarray}
i\left( \hat{k}_{1}\frac{\partial }{\partial z}+\frac{1}{v_{1}}\frac{%
\partial }{\partial t}\right) \tilde{E}_{1}-\frac{D_{1}}{2}\frac{\partial
^{2}}{\partial t^{2}}\tilde{E}_{1} &=&-\frac{2\pi \omega _{1}^{2}\mu (\omega
_{1})}{c^{2}k_{1}}\tilde{P}_{NL}(\omega _{1})\exp (-ik_{1}z)  \label{eq1a} \\
i\left( \hat{k}_{2}\frac{\partial }{\partial z}+\frac{1}{v_{2}}\frac{%
\partial }{\partial t}\right) \tilde{E}_{2}-\frac{D_{2}}{2}\frac{\partial
^{2}}{\partial t^{2}}\tilde{E}_{2} &=&-\frac{2\pi \omega _{2}^{2}\mu (\omega
_{2})}{c^{2}k_{2}}\tilde{P}_{NL}(\omega _{2})\exp (-ik_{2}z)  \label{eq1b}
\end{eqnarray}%
where $\omega _{1}=\omega _{p},\omega _{2}=\omega _{s}=2\omega _{p}$, $%
k_{j}^{2}=(\omega _{j}/c)^{2}\varepsilon (\omega _{j})\mu (\omega _{j})$, $%
\hat{k}_{j}$ is sign of square root of $k_{j}^{2}$, and

\begin{eqnarray*}
\tilde{P}_{NL}(\omega _{1}) &=&\chi _{2}(\omega _{1};\omega _{2},-\omega
_{1})\tilde{E}_{2}\tilde{E}_{1}^{\ast }\exp [iz(k_{2}-k_{1})], \\
\tilde{P}_{NL}(\omega _{2}) &=&\chi _{2}(\omega _{2};\omega _{1},\omega _{1})%
\tilde{E}_{1}^{2}\exp [iz(k_{1}+k_{1})].
\end{eqnarray*}%
\bigskip where $D_{1,2}$ are dispersion coefficients at pump and signal wave
frequencies \cite{R3,R4}. Here $\tilde{E}_{1}$ is slowly varying envelope of
the pump (fundamental frequency) wave, and $\tilde{E}_{2}$ is the slowly
varying envelope of the second harmonic wave.

As we have in equations (\ref{eq1a},\ref{eq1b}) ratio $(\omega _{j}/c)\sqrt{%
\mu (\omega _{j})/\varepsilon (\omega _{j})}$, the coefficients in right
part of equations (\ref{eq1a},\ref{eq1b}) are not depending on signs of the
dielectric permittivity and magnetic permeability. Nevertheless direction of
the wave propagation is defined by signs $\hat{k}_{j}$.

It should be pointed that we consider the collinear propagation and so the
collinear phase mismatch. The condition $\Delta k=0$ means that vector $\vec{%
k}_{2}$ of pump wave is collinear and equal to vector $2\vec{k}_{1}$. The
configuration of these vectors defines the energy flow directions. If the
nonlinear medium is characterized by positive refraction index, then vectors 
$\vec{k}_{1,2}$\ and Pointing vectors have same directions. Opposite figure
take place if the nonlinear medium is characterized by negative refraction
index. In this case waves with frequencies from frequency region of negative
refraction index are directed according to zero phase mismatch condition,
but the Pointing vectors are contra-directed \cite{R25,R5} (see Fig.1).

We will consider the SHG under assumption that frequency of pump wave belong
the frequency region of negative refraction index, while the frequency of
the signal wave lies in frequency region positive refraction index.

Let us introduce the normalized variables%
\begin{equation}
\zeta =z/L,\tau =(t+z/v_{1})/t_{p},\quad \tilde{E}_{1}=A_{10}e_{1},\tilde{E}%
_{2}=A_{20}e_{2}\exp (iz\Delta k),  \label{eq2}
\end{equation}

\begin{eqnarray}
A_{10} &=&c(2\pi \omega _{1}L\chi _{2}(\omega _{1}))^{-1}\sqrt{\varepsilon
(\omega _{1})/\mu (\omega _{1})},  \label{eq3} \\
A_{20} &=&c(2\pi \omega _{2}L\chi _{2}(\omega _{2}))^{-1}\sqrt{\varepsilon
(\omega _{2})/\mu (\omega _{2})}, \\
\chi _{1} &=&\chi ^{(2)}(\omega _{1};2\omega _{1},-\omega _{1}),\quad \chi
_{2}=\chi ^{(2)}(2\omega _{1};\omega _{1},\omega _{1}), \\
\mu _{1,3} &=&\mu (\omega _{1,3}),\quad \varepsilon _{1,3}=\varepsilon
(\omega _{1,3}),
\end{eqnarray}%
where $\Delta k=k_{2}-2k_{1}$. Then SHG is described by the following
normalized system of the equations%
\begin{equation}
i\frac{\partial }{\partial \zeta }e_{1}+\frac{\sigma }{2}\frac{\partial ^{2}%
}{\partial \tau ^{2}}e_{1}-e_{1}^{\ast }e_{2}=0  \label{eq4a}
\end{equation}%
\begin{equation}
i(\frac{\partial }{\partial \zeta }+\delta \frac{\partial }{\partial \tau }%
)e_{2}-\frac{\beta }{2}\frac{\partial ^{2}}{\partial \tau ^{2}}e_{2}-\Delta
e_{2}+\frac{\theta }{2}e_{1}^{2}=0  \label{eq4b}
\end{equation}%
where

\[
\delta =Lt_{p}^{-1}(v_{1}^{-1}+v_{2}^{-1})>0,\quad \theta =\left( \frac{%
2\chi _{1}^{(2)}}{\chi _{2}^{(2)}}\right) ^{2}\sqrt{\frac{\mu (\omega
_{1})\varepsilon (2\omega _{1})}{\mu (2\omega _{1})\varepsilon (\omega _{1})}%
}\approx 1 
\]%
\[
\sigma =\mathrm{sgn}D_{1},\quad \beta =D_{2}/|D_{1}|,\quad
L=t_{p}^{2}/|D_{1}|. 
\]

Parameter $\delta $ takes account of the walk-off effect for pump and
harmonic pulses that is due to the difference of the group velocities
directions for interaction waves. This parameter can not be zero, as in the
case of positive refractive index medium.

\section{Integral of motion - the Manley-Rowe relation}

By using the equations (\ref{eq4a},\ref{eq4b}) the following expression can
be obtained%
\[
i\frac{\partial }{\partial \zeta }|e_{1}|^{2}+\frac{\sigma }{2}\left(
e_{1}^{\ast }\frac{\partial ^{2}e_{1}}{\partial \tau ^{2}}-e_{1}\frac{%
\partial ^{2}e_{1}^{\ast }}{\partial \tau ^{2}}\right) -e_{1}^{\ast
2}e_{2}+e_{2}^{\ast }e_{1}^{2}=0, 
\]%
\[
i(\frac{\partial }{\partial \zeta }+\delta \frac{\partial }{\partial \tau }%
)|e_{2}|^{2}-\frac{\beta }{2}\left( e_{2}^{\ast }\frac{\partial ^{2}e_{2}}{%
\partial \tau ^{2}}-e_{2}\frac{\partial ^{2}e_{2}^{\ast }}{\partial \tau ^{2}%
}\right) -\frac{1}{2}\left( e_{1}^{\ast 2}e_{2}-e_{2}^{\ast
}e_{1}^{2}\right) =0. 
\]%
From these equations it follows that%
\begin{eqnarray*}
0 &=&i\frac{\partial }{\partial \zeta }\left( |e_{2}|^{2}-\frac{1}{2}%
|e_{1}|^{2}\right) +i\delta \frac{\partial }{\partial \tau }|e_{2}|^{2}- \\
&&-\frac{\beta }{2}\frac{\partial }{\partial \tau }\left( e_{2}^{\ast }\frac{%
\partial e_{2}}{\partial \tau }-e_{2}\frac{\partial e_{2}^{\ast }}{\partial
\tau }\right) -\frac{\sigma }{4}\frac{\partial }{\partial \tau }\left(
e_{1}^{\ast }\frac{\partial e_{1}}{\partial \tau }-e_{1}\frac{\partial
e_{1}^{\ast }}{\partial \tau }\right) .
\end{eqnarray*}%
If we take into account the boundary condition%
\[
|e_{1,2}|^{2}\rightarrow |e_{10,20}|^{2}=\mathrm{const}, 
\]%
or%
\[
|e_{1,2}|^{2}\rightarrow 0 
\]%
at $|\tau |\rightarrow \pm \infty $, then 
\[
\dint\limits_{-\infty }^{\infty }\left( |e_{2}|^{2}-\frac{1}{2}%
|e_{1}|^{2}\right) d\tau =\mathrm{const.} 
\]

In the case of the SHG in continuum wave that results in the Manley-Rowe
relation \cite{R6}%
\[
|e_{2}|^{2}-\frac{1}{2}|e_{1}|^{2}=\mathrm{const.} 
\]

When SGH takes place in positive refractive index medium the minus in this
expression is replaced to sign plus. It corresponds to the propagation of
the both interacting waves in one direction.

\section{Real form of the equations}

It is suitable to introduce the real variables for the interacting waves%
\begin{equation}
e_{1}=a\exp (i\varphi _{1}),\quad e_{2}=b\exp (i\varphi _{2}).  \label{eq6}
\end{equation}%
Substitution of these expressions into (\ref{eq4a},\ref{eq4b}) leads to

\begin{eqnarray}
\frac{\partial a}{\partial \zeta }+\frac{\sigma }{2}\left( 2\frac{\partial a%
}{\partial \tau }\frac{\partial \varphi _{1}}{\partial \tau }+a\frac{%
\partial ^{2}\varphi _{1}}{\partial \tau ^{2}}\right) &=&ab\sin \Phi ,
\label{eq7a} \\
\frac{\partial b}{\partial \zeta }+\delta \frac{\partial b}{\partial \tau }-%
\frac{\beta }{2}\left( 2\frac{\partial b}{\partial \tau }\frac{\partial
\varphi _{2}}{\partial \tau }+b\frac{\partial ^{2}\varphi _{2}}{\partial
\tau ^{2}}\right) &=&\frac{1}{2}a^{2}\sin \Phi ,  \label{eq7b} \\
a\frac{\partial \varphi _{1}}{\partial \zeta }-\frac{\sigma }{2}\left( \frac{%
\partial ^{2}a}{\partial \tau ^{2}}-a\frac{\partial \varphi _{1}}{\partial
\tau }\frac{\partial \varphi _{1}}{\partial \tau }\right) &=&-ab\cos \Phi ,
\label{eq7c} \\
a\left( \frac{\partial \varphi _{2}}{\partial \zeta }+\delta \frac{\partial
\varphi _{2}}{\partial \tau }\right) +\frac{\beta }{2}\left( \frac{\partial
^{2}b}{\partial \tau ^{2}}+b\frac{\partial \varphi _{2}}{\partial \tau }%
\frac{\partial \varphi _{2}}{\partial \tau }\right) +\Delta b &=&\frac{1}{2}%
a^{2}\cos \Phi ,  \label{eq7d}
\end{eqnarray}%
where $\Phi =\varphi _{2}-2\varphi _{1}$. The case of continuum wave has
been considered in \cite{R5,R6}, where principal difference between spatial
distribution of the pump and harmonic fields in the NRI medium with respect
to positive refraction index medium was pointed.

\section{Steady state solutions}

Let us consider solutions without chirp, i.e., ones with $\partial
^{2}\varphi _{1,.2}/\partial \tau ^{2}=0 $. Hence, we have $\partial \varphi
_{1,.2}/\partial \tau =\Omega _{1,2}$. The mismatch condition $\Phi =0$,
which is assumed additionally, leads to $\Omega _{2}=2\Omega _{1}=2\Omega $.
Furthermore, we assume that $\partial \varphi _{1,.2}/\partial \zeta =\kappa
_{1,2}$., thus we suppose the phase functions evolve as%
\begin{equation}
\varphi _{1}=\kappa \zeta +\Omega \tau ,\quad \varphi _{2}=2\kappa \zeta
+2\Omega \tau .  \label{eq8}
\end{equation}

The amplitude equations (\ref{eq7a}) and (\ref{eq7b}) take the following form%
\begin{eqnarray}
\frac{\partial a}{\partial \zeta }+\sigma \Omega \frac{\partial a}{\partial
\tau } &=&0,  \label{eq9a} \\
\frac{\partial b}{\partial \zeta }+(\delta -2\beta \Omega )\frac{\partial b}{%
\partial \tau } &=&0.  \label{eq9b}
\end{eqnarray}%
The steady state waves for both frequencies must propagate as single one.
Thus, from (\ref{eq9a}, \ref{eq9b}) we get the amplitudes as function of
single variable $y=\tau -\zeta /V$ 
\[
a=a(\tau -\zeta /V),\quad b=b(\tau -\zeta /V) 
\]%
with%
\[
V^{-1}=\sigma \Omega =\delta -2\beta \Omega . 
\]%
This expression defines the parameter "instant frequency" \ $\Omega =\delta
/(\sigma +2\beta )$.

The phase equations (\ref{eq7c}) and (\ref{eq7d}) take the following form%
\begin{equation}
\frac{\partial ^{2}a}{\partial \tau ^{2}}-(2\sigma \kappa +\Omega
^{2})a=2\sigma ab,  \label{eq13c}
\end{equation}%
\begin{equation}
\frac{\partial ^{2}b}{\partial \tau ^{2}}+\frac{2}{\beta }\left( \Delta
+2\kappa +2\delta \Omega -2\beta \Omega ^{2}\right) b=\frac{1}{\beta }a^{2}.
\label{eq13d}
\end{equation}

Let us suppose that $b=fa$. From (\ref{eq13c}) and (\ref{eq13d}) we get two
equations for one function $a=a(y)$ of one variable $y=\tau -\zeta /V $ :%
\begin{eqnarray}
\frac{\partial ^{2}a}{\partial y^{2}}-(2\sigma \kappa +\Omega ^{2})a
&=&2\sigma fa^{2},  \label{eq14a} \\
\frac{\partial ^{2}a}{\partial y^{2}}+\frac{2}{\beta }\left( \Delta +2\kappa
+2\delta \Omega -2\beta \Omega ^{2}\right) a &=&\frac{1}{\beta f}a^{2}
\label{eq14b}
\end{eqnarray}%
These equations would define single function $a$, if the coefficients before
equal orders of $a$ are same, i.e., $f^{2}=\sigma /2\beta >0$ and%
\begin{equation}
-(2/\beta )\left( \Delta +2\kappa +2\delta \Omega -2\beta \Omega ^{2}\right)
=(2\sigma \kappa +\Omega ^{2}).  \label{eq15}
\end{equation}%
It should be remarked that condition $\sigma ^{2}=1$ was used. This
expression defines the parameter (effective wave number) $\kappa $ 
\[
\kappa =\frac{3\beta \Omega ^{2}-4\delta \Omega -2\Delta }{2\sigma (2\sigma
+\beta )}. 
\]

From equation for function $a(y)$ (\ref{eq14a}) we can obtain%
\begin{equation}
\left( \partial a/\partial y\right) ^{2}=(2\sigma \kappa +\Omega
^{2})a^{2}+(4\sigma f/3)a^{3}+\mathrm{const.}  \label{eq16}
\end{equation}

If $\sigma =+1$ , then dispersion parameter of second harmonic $\beta $ must
be positive too. Else, if $\sigma =-1$, then parameter $\beta $ must be
negative.

Let be $\sigma =-1.$ In this case we have $f=(1/2|\beta |)^{1/2}$, and the
equation (\ref{eq16}) can be rewritten as%
\begin{equation}
\left( \partial a/\partial y\right) ^{2}=(\Omega ^{2}-2\kappa
)a^{2}-(4f/3)a^{3}+\mathrm{const.}  \label{eq17}
\end{equation}%
Let us introduce $p=(\Omega ^{2}-2\kappa )$ and $w(y)=(4f/3p)a(y)$. Consider
the boundary condition as following: $a\rightarrow a_{0}$, $\partial
a/\partial y\rightarrow 0$ at $\tau \rightarrow \pm \infty $. When $a_{0}=0$%
, we will get the solitary wave on zero background. The constant of
integrating is defined now by following expression $\mathrm{const}%
=-p(w_{0}^{2}-w_{0}^{3})$, where $w_{0}=(4f/3p)a_{0}$ . From equation (\ref%
{eq17}) the equation for $w(y)$ follows%
\begin{equation}
\left( \partial w/\partial y\right) ^{2}=p(w-w_{0})\left[ \Delta
^{2}-(w-\gamma _{0})^{2}\right] ,  \label{eq18}
\end{equation}%
where 
\[
\Delta ^{2}=(1-w_{0})(1+3w_{0})/4,\quad \gamma _{0}=(1-w_{0})/2. 
\]

If we introduce new variable $u$ through the formula $w=w_{0}+u^{-2}$\ ,
then the equation (\ref{eq18}) transforms in to%
\begin{equation}
\left( \partial u/\partial y\right) ^{2}=(p/4)\left[ \Delta
^{2}u^{4}-(1-\gamma _{1}u^{2})^{2}\right] ,  \label{eq19}
\end{equation}%
\bigskip where $\gamma _{1}=(1-3w_{0})/2$ . Parameter $\Delta ^{2}$ is
positive at $0\leq w_{0}<1$.

\subsection{Bright soliton solution}

The solution of this kind corresponds with choosing $w_{0}=0,\partial
a/\partial y\rightarrow 0$ at $\tau \rightarrow \pm \infty $. In this case
the parameters in (\ref{eq19}) are $\Delta ^{2}=1/4$ and $\gamma _{1}=1/2$.
The equation (\ref{eq19}) transforms into following one%
\begin{equation}
\left( \partial u/\partial y\right) ^{2}=(p/4)(u^{2}-1),  \label{eq20}
\end{equation}%
Put $u=\cosh \vartheta $, then from this equation we get $\left( \partial
\vartheta /\partial y\right) ^{2}=(p/4\dot{)}>0$. Hence, solution of (\ref%
{eq20}) is 
\begin{equation}
u(y)=\cosh \left[ p^{1/2}(y-y_{0})/2\right] ,  \label{eq21}
\end{equation}%
where $y_{0}$\ is constant of integrating.\ Steady state solution of
equations (\ref{eq7a}- \ref{eq7d}) in the case of $\sigma =-1$ corresponding
with \emph{bright soliton} is 
\begin{eqnarray}
a(y) &=&\frac{(3p/4)\sqrt{2|\beta |}}{\cosh ^{2}[\sqrt{p}(y-y_{0})/2]},
\label{eq22a} \\
b(y) &=&\frac{(3p/4)}{\cosh ^{2}[\sqrt{p}(y-y_{0})/2]}.  \label{eq22b}
\end{eqnarray}

If \ $p=-|p|<0$, equation (\ref{eq20}) has periodical solution%
\[
u(y)=\cos \left[ \sqrt{|p|}(y-y_{0})/2\right] , 
\]
which leads to singular solution of the system (\ref{eq7a}- \ref{eq7d}).

If $\sigma =+1$ , then $f=(1/2\beta )^{1/2}$\ .\ The equation (\ref{eq16})
takes the form 
\begin{equation}
\left( \partial a/\partial y\right) ^{2}=(\Omega ^{2}+2\kappa
)a^{2}+(4f/3)a^{3}+\mathrm{const.}  \label{eqA}
\end{equation}%
By introducing new function $\widetilde{w}(y)=-(4f/3p)a(y)$\ from (\ref{eqA}%
) the equation for $\widetilde{w}(y)$ follows%
\[
\left( \partial \widetilde{w}/\partial y\right) ^{2}=\widetilde{p}(%
\widetilde{w}-\widetilde{w}_{0})\left[ \widetilde{\Delta }^{2}-(\widetilde{w}%
-\widetilde{\gamma }_{0})^{2}\right] , 
\]%
\ \ \ where $\widetilde{p}=(\Omega ^{2}+2\kappa )$ , $\widetilde{w}%
_{0}=-(4f/3p)a_{0}$\ , and%
\[
\widetilde{\Delta }^{2}=(1-\widetilde{w}_{0})(1+3\widetilde{w}_{0})/4,\quad 
\widetilde{\gamma }_{0}=(1-\widetilde{w}_{0})/2. 
\]%
Hence, this case reduced to the one considered above.

\subsection{Cnoidal wave solution}

The solution of this kind corresponds with choosing $0<w_{0}<1$. If \ $%
0<w_{0}<1/3$\ then we have $\Delta >\gamma _{1}>0$. Equation (\ref{eq19})
can be written as%
\begin{equation}
\left( \partial u/\partial y\right) ^{2}=(p/4)\left[ (\Delta +\gamma
_{1})u^{2}-1\right] \left[ (\Delta -\gamma _{1})u^{2}+1\right] .
\label{eq23}
\end{equation}%
Introduce the new variable $\phi $ according to expression $\cosh \phi =u%
\sqrt{\Delta +\gamma _{1}}$. It leads to%
\begin{equation}
\left( \partial \phi /\partial y\right) ^{2}=(p/4)(\Delta +\gamma _{1})\left[
1+m^{2}\cosh ^{2}\phi \right]  \label{eq24}
\end{equation}%
where $m^{2}=(\Delta -\gamma _{1})(\Delta +\gamma _{1})^{-1}$. Substitution $%
\phi =i\psi $\ into this equation leads to%
\[
\left( \partial \psi /\partial y\right) ^{2}=-(p/4)(\Delta +\gamma
_{1})(1+m^{2})\left[ 1-\widetilde{m}^{2}\sin ^{2}\psi \right] . 
\]%
It is equation for the Jacobi elliptic functions. Hence, we can write the
solution by following formula%
\begin{equation}
\sin \psi =\mathrm{sn}(iz,\widetilde{m}),  \label{eq25}
\end{equation}%
where $z=\sqrt{p(\Delta +\gamma _{1})(1+m^{2})}(y-y_{0})/2$, $\widetilde{m}%
^{2}=m^{2}(1+m^{2})^{-1}$. By using the properties of the trigonometric
functions and Jacobi elliptic functions, in particular, $\mathrm{sn}(iz,k)=i%
\mathrm{sc}(z,k^{\prime })$,$k^{\prime 2}+k^{2}=1$ , and $\mathrm{sn}%
^{2}(z,k)+\mathrm{cn}^{2}(z,k)=1$, we can write 
\[
\sinh \phi =-\frac{\mathrm{sn}(z,\widetilde{m}^{\prime })}{\mathrm{cn}(z,%
\widetilde{m}^{\prime })}. 
\]%
Now, we have 
\begin{equation}
u^{2}(y)=\frac{1}{(\Delta +\gamma _{1})\mathrm{cn}^{2}(z,\widetilde{m}%
^{\prime })}.  \label{eq26}
\end{equation}%
Thus, the solution of the equation (\ref{eq18}) represents by formula%
\begin{equation}
w(y)=w_{0}+(\Delta +\gamma _{1})\mathrm{cn}^{2}(z,\widetilde{m}^{\prime }).
\label{eq27}
\end{equation}%
By using definitions of the parameter $\widetilde{m}$\ and $m^{2}$\ one can
find that%
\[
\widetilde{m}^{\prime }=\left[ \frac{1}{2}\left( 1+\frac{\gamma _{1}}{\Delta 
}\right) \right] ^{1/2},\quad z=\left( \frac{2p\Delta }{\Delta +\gamma _{1}}%
\right) ^{1/2}(y-y_{0}). 
\]

At interval $0<w_{0}<1/3$ module of Jacobi elliptic functions is not equal
to unit, hence only cnoidal wave exists at these values of the pedestal
amplitude $w_{0}$.

The solution of the equation (\ref{eq19})\ under condition that $1/3<w_{0}<1$%
\ can be found in a similar way. If $1/3<w_{0}<2/3$\ this solution
represents by formula%
\begin{equation}
w(y)=w_{0}+(\Delta -|\gamma _{1}|)\mathrm{cn}^{2}(z_{1},m_{1}),  \label{eq30}
\end{equation}
\ where 
\[
m_{1}=\left[ \frac{1}{2}\left( 1-\frac{|\gamma _{1}|}{\Delta }\right) \right]
^{1/2},\quad z_{1}=\left( \frac{p\Delta }{2}\right) ^{1/2}(y-y_{0}). 
\]

If \ $2/3<w_{0}<1$ we can find \ 
\begin{equation}
w(y)=w_{0}+(\Delta +|\gamma _{1}|)\frac{\mathrm{cn}^{2}(z_{2},m_{2})}{%
\mathrm{sn}^{2}(z_{2},m_{2})},  \label{eq31}
\end{equation}%
where 
\[
m_{2}=\left[ \frac{2\Delta }{\Delta +|\gamma _{1}|}\right] ^{1/2},\quad
z_{2}=\frac{1}{2}\left[ |p|(\Delta +|\gamma _{1}|)\right] ^{1/2}(y-y_{0}). 
\]%
This solution describes the cnoidal wave, however amplitude of this wave
periodically attains the infinite value. Hence, this solution is not
acceptable for considered there physical situation. If \ $w_{0}>1$ equation (%
\ref{eq19}) has only singular (unlimited) solutions.

\subsection{Dark soliton solution}

The quadratic dark soliton has been found in \cite{R40,R41} . The similar
solution can be found in the NRI medium. Let consider the equation (\ref%
{eq18}). Substitution in $w=w_{0}+q$ results in the following equation%
\begin{equation}
\left( \partial q/\partial y\right) ^{2}=pq\left[ \Delta ^{2}-(q-\gamma
_{1})^{2}\right] .  \label{eq32}
\end{equation}%
By following \cite{R41}\ \ let us find the solution in the form $q=A/\cosh
\alpha y$. If substitute that in (\ref{eq32}), and equating the coefficients
of various powers of\ \ $\cosh \alpha y$\ \ in this equation to zero, we
obtain the following system of relations%
\begin{equation}
p(\Delta ^{2}-\gamma _{1}^{2})=0,\quad 2p\gamma _{1}A=4\alpha ^{2}A,\quad
A^{2}=4\alpha A.  \label{eq33}
\end{equation}%
From these relations it follows that $\alpha ^{2}=p\gamma _{1}/2$, $%
A=4\alpha $, and $\Delta ^{2}-\gamma _{1}^{2}=0$. By using the definition of 
$\Delta ^{2}$\ and $\gamma _{1}$ we can find that equation $\Delta
^{2}=\gamma _{1}^{2}$ will be held at $w_{0}=0$ and $w_{0}=2/3$. In the past
case $\gamma _{1}=-1/2$. From relation $\alpha ^{2}=p\gamma _{1}/2$\ it
follows that $p$ must be negative, and $\alpha =\pm (|p|/4)^{1/2}$. Hence,
we have $A=\pm 2(|p|)^{1/2}$. Thus,%
\[
q=\frac{\pm 2\sqrt{|p|}}{\cosh ^{2}[\sqrt{|p|}y/2]}. 
\]%
The solution of equations (\ref{eq7a}- \ref{eq7d}) in this case is%
\begin{equation}
a(y)=(3|p|/4)\sqrt{2|\beta |}\left( 2/3\pm 2\sqrt{|p|}\sec \mathrm{h}^{2}[%
\sqrt{|p|}y/2]\right)  \label{eq34a}
\end{equation}%
\begin{equation}
b(y)=(3|p|/4)\left( 2/3\pm 2\sqrt{|p|}\sec \mathrm{h}^{2}[\sqrt{|p|}%
y/2]\right)  \label{eq34b}
\end{equation}%
These expressions describe the quadratic dark soliton propagating in
nonlinear medium with negative refractive index.

\subsection{Two-hump soliton}

Up to this point the solutions of equations (\ref{eq13c}) and (\ref{eq13d})
have been found under constraint $b=fa$. It is known \cite{R42} that for
positive refractive index medium there are solutions where this constraint
is absent. These solutions were named by \emph{simultons} in \cite{R33}. We
can show that the similar solutions exist \ in NRIM too. Follow \cite{R42}
we assume that solution of the (\ref{eq13c}) and (\ref{eq13d}) has the form 
\begin{eqnarray}
a(y) &=&A\tanh \alpha y\sec \mathrm{h}\alpha y,  \label{eq35a} \\
b(y) &=&B\sec \mathrm{h}^{2}\alpha y,  \label{eq35b}
\end{eqnarray}%
where $A$, $B$ and $\alpha $ are unknown parameters. Substitution of this
anzats\ in (\ref{eq13c}) and (\ref{eq13d}), and grouping the terms with the
equal powers of\ \ $\cosh \alpha y$\ \ and $\tanh \alpha y\sec \mathrm{h}%
\alpha y$\ in these equations to zero, provides the following system of
algebraic relations%
\begin{eqnarray}
\alpha ^{2}A-p_{1}A &=&0,  \label{eq36a} \\
2\sigma AB+6\alpha ^{2}A &=&0,  \label{eq36b} \\
A^{2}/\beta -6\alpha ^{2}A &=&0,  \label{eq36c} \\
-A^{2}/\beta +4\alpha ^{2}B &=&p_{2},  \label{eq36d}
\end{eqnarray}%
where $(2\sigma \kappa +\Omega ^{2})=p_{1}$ and $(2/\beta )\left( \Delta
+2\kappa +2\delta \Omega -2\beta \Omega ^{2}\right) =p_{2}$. It leads to%
\begin{eqnarray*}
\alpha ^{2} &=&p_{1}>0,\quad A^{2}=6\beta \alpha ^{2}B, \\
B &=&p_{2}/2\alpha ^{2},\quad B=-3\alpha ^{2}\sigma .
\end{eqnarray*}%
From these expression we obtain $A^{2}=-18\beta \alpha ^{4}\sigma $ . It
means that if $\sigma =-1$ then $\beta >0$ . Else, we can chouse $\sigma =1$%
, then $\beta <0$\ and $p_{2}=-|p_{2}|<0$\ .

Let be $\sigma =-1$. In this case we can found%
\begin{equation}
\alpha ^{2}=p_{1},\quad B=3p_{1},\quad A=3\sqrt{2\beta }p_{1},  \label{eq37a}
\end{equation}%
and relation%
\begin{equation}
p_{2}=6p_{1}^{2}.  \label{eq37b}
\end{equation}
From (\ref{eq37b}) it follows%
\begin{equation}
\left( \Delta +2\kappa +2\delta \Omega -2\beta \Omega ^{2}\right) =3\beta
\left( 2\kappa -\Omega ^{2}\right) ^{2}.  \label{eq37c}
\end{equation}

The parameter $\Omega $ has been defined as\ $\Omega =\delta /(2\beta -1)$,
hence relation (\ref{eq37c}) defines parameter through phase mismatch $%
\Delta $. From (\ref{eq37c}) it follows%
\begin{equation}
2\kappa ^{(\pm )}=\frac{1+6\beta \Omega ^{2}\pm \sqrt{1+3\beta \left( \Delta
+2(3\beta -1)\Omega ^{2}+9\beta \Omega ^{4}\right) }}{6\beta }.  \label{eq38}
\end{equation}

Thus, the two-hump soliton has the following form%
\begin{eqnarray}
a(y) &=&3p_{1}\sqrt{2\beta }\tanh \left( \sqrt{p_{1}}y\right) \sec \mathrm{h}%
\left( \sqrt{p_{1}}y\right) ,  \label{eq39a} \\
b(y) &=&3p_{1}\sec \mathrm{h}^{2}\left( \sqrt{p_{1}}y\right) .  \label{eq39b}
\end{eqnarray}%
For example, let us assume that $\beta =1$, then $\Delta =-\delta ^{2}$, $%
2\kappa ^{(\pm )}=\delta ^{2}\pm 1/3$\ and $p_{1}^{(\pm )}=\pm 1/3$. Only $%
p_{1}^{(+)}=1/3$\ is acceptable, hence, we get the two-hump solution of the
system (\ref{eq13c}) and (\ref{eq13d}) 
\begin{eqnarray}
a(y) &=&\sqrt{2}\tanh \left( y/\sqrt{3}\right) \sec \mathrm{h}\left( y/\sqrt{%
3}\right) ,  \label{eq40a} \\
b(y) &=&\sec \mathrm{h}^{2}\left( y/\sqrt{3}\right) .  \label{eq40b}
\end{eqnarray}

For the case of the positive refractive index medium there is the solution
describing the perimetrically connected waves of the pump and second
harmonic. Analogue of that exists in the NRI medium too.

\section{Conclusion}

We considered the steady state propagation of the coupled pair of waves,
i.e., the pump wave at the frequency $\omega _{0}$ and wave at the second
harmonic frequency, in the quadratic nonlinear NRIM. In this case the group
velocities of pump and harmonic pulses are directed in opposite directions 
\cite{R25, R26, R5, R6}. We are foud that under some conditions the self
trapping of the interacting wave packets takes place. Two frequency wave
packet (simulton) is generated from initial pulses alike to quadratic
soliton of positive refractive medium (PRIM).

To do comparison of the solitons in quadratic NRIM with solitons in PRIM it
is suitable to represent the appropriat expressions for second case. The
system of equations describing the quadratic solitons propagation in PRI
medium can be written as 
\begin{equation}
i\frac{\partial }{\partial \zeta }e_{1}-\frac{\sigma }{2}\frac{\partial ^{2}%
}{\partial \tau ^{2}}e_{1}+e_{1}^{\ast }e_{2}=0  \label{eq41a}
\end{equation}%
\begin{equation}
i(\frac{\partial }{\partial \zeta }+\delta \frac{\partial }{\partial \tau }%
)e_{2}-\frac{\beta }{2}\frac{\partial ^{2}}{\partial \tau ^{2}}e_{2}-\Delta
e_{2}+\frac{1}{2}e_{1}^{2}=0  \label{eq41b}
\end{equation}%
where $\delta
=Lt_{p}^{-1}(v_{1}^{-1}-v_{2}^{-1})=t_{p}|D_{1}|^{-1}(v_{1}^{-1}-v_{2}^{-1})$%
. In NRIM parameter $\delta $ is always positive. Hence, there is no
solutions corresponding with solitons of PRIM which have $\delta =0$

Velocities of the bright solitons in NRIM and PRIM are differed. For NRIM
all steady state pulses move with velocity

\begin{equation}
\frac{1}{V_{s}}=\frac{1}{\sigma +2\beta }\left( \frac{\sigma }{v_{2}}-\frac{%
2\beta }{v_{1}}\right) =\frac{1}{D_{1}+2D_{2}}\left( \frac{D_{1}}{v_{2}}-%
\frac{2D_{2}}{v_{1}}\right) .  \label{eq42a}
\end{equation}%
Whereas in PRI medium bright soliton has the velocity 
\begin{equation}
\frac{1}{V_{s}}=\frac{1}{2\beta -\sigma }\left( \frac{2\beta }{v_{1}}-\frac{%
\sigma }{v_{2}}\right) =\frac{1}{2D_{3}-D_{1}}\left( \frac{2D_{2}}{v_{1}}-%
\frac{D_{1}}{v_{2}}\right) .  \label{eq42b}
\end{equation}

In the plane $\left( \delta ,\Delta \right) $ the existence regions of the
steady state wave are differed also. For example, if one put dispersion
parameters as follows $\sigma =\beta =-1$ in the case of NRI medium from (%
\ref{eq22a}-\ref{eq22b}) the expressions for envelopes of bright soliton
result in 
\begin{equation}
a(y)=\frac{\sqrt{2}(3\Delta -\delta ^{2})}{6\cosh ^{2}[\sqrt{(3\Delta
-\delta ^{2})/2}(y-y_{0})/3]},  \label{eq43a}
\end{equation}%
\begin{equation}
b(y)=\frac{(3\Delta -\delta ^{2})}{6\cosh ^{2}[\sqrt{(3\Delta -\delta ^{2})/2%
}(y-y_{0})/3]}.  \label{eq43b}
\end{equation}%
This solution exists under condition $\Delta >\delta ^{2}/3$ . The solution
of the equation (\ref{eq41a}) and (\ref{eq41b}) yields expression for
quadratic soliton in PRIM 
\begin{equation}
a(y)=\frac{3\sqrt{2}(\Delta +\delta ^{2})}{2\cosh ^{2}[\sqrt{(\Delta +\delta
^{2})/2}(y-y_{0})]},  \label{eq44a}
\end{equation}%
\begin{equation}
b(y)=\frac{3(\Delta +\delta ^{2})}{2\cosh ^{2}[\sqrt{(\Delta +\delta ^{2})/2}%
(y-y_{0})]}.  \label{eq44b}
\end{equation}%
In this case the solution exists under condition $\Delta +\delta ^{2}>0$ .
Thus the NRIM solitons and PRIM solitons exist in different regions of
parameters plane\ \ $\left( \delta ,\Delta \right) $. Stability conditions
we plane to consider in future investigation. It is possible that the
stability regions will be different too.

\section*{Acknowledgment}

It is a pleasure for the authors to thank Dr. A.M. Basharov and Dr. S.O.
Elyutin for very useful discussions. A.I. Maimistov is grateful to the
Department of Mathematics, University of Arizona for hospitality and
support. This research was supported in part by RFBR (Grant No 06-02-16406),
by NSF (Grant DMS-0509589) and by State Arizona (Proposition 301)


\bigskip

\begin{thebibliography}{99}
\bibitem{M1} Mandelschtam L.I., \textit{Zh.Exp.Teor.Fiz}. \textbf{15}.
475-478 (1945).

\bibitem{M2} Veselago V.G., \textit{Sov. Phys. Solid State} \textbf{8}, 2854
(1967)

\bibitem{M3} Pendry J. B., Holden A. J., Stewart W. J., Youngs I., \textit{%
Phys.Rev.Lett.} \textbf{76}. 4773-4776 (1996).

\bibitem{M4} Smith D. R., Vier D. C., Kroll N., Schultz S. \textit{%
Appl.Phys.Lett.} \textbf{77}. 2246-2248 (2000).

\bibitem{M5} Smith D. R., Kroll N. \textit{Phys.Rev.Lett.} \textbf{85}%
,2933-2936 (2000).

\bibitem{M6} Pendry J.B. \textit{Phys.Rev.Lett.} \textbf{85}, 3966-3969
(2000).

\bibitem{M7} Richard W. Ziolkowski, Ehud Heyman. \textit{Phys.Rev.} \textbf{%
B64}. 056625 (2001).

\bibitem{M8} Smith D.R., Padilla W.J., Vier D.C., Nemat-Nasser S.C.,
SchultzS. \textit{Phys.Rev.Lett.} \textbf{84}. 4184-4187 (2000).

\bibitem{M9} Shelby R.A., Smith D.R., Schultz S. \textit{Science} \textbf{292%
}, 77-79 (2001).

\bibitem{M10} Parazzoli C.G., Greegor R.B., Li K., Koltenbah
B.E.C.,Tanielian M. \textit{Phys. Rev. Lett.} \textbf{90}, 107401 (2003).

\bibitem{M11} Houck A.A., Brock J.B., Chuang I.L. \textit{Phys. Rev. Lett.} 
\textbf{90}, 137401 (2003).

\bibitem{M12} Shalaev V. M., Cai W., Chettiar U., Yuan H.-K., Sarychev A.
K., Drachev V. P., Kildishev A. V. \emph{Negative Index of Refraction in
Optical Metamaterials}. arXiv. phys-ics/0504091.

\bibitem{M13} Drachev V.P., Cai W., Chettiar U., Yuan H.-K., Sarychev A.K.,
Kildishev A.V., Klimeck G., Shalaev V.M. \textit{Laser Phys. Lett}. \textbf{3%
}, 49-55 (2006).

\bibitem{M14} Shalaev V. M., Cai W., Chettiar U., Yuan H.-K., Sarychev A.
K., Drachev V. P., Kildishev A. V. \textit{Opt. Lett}. \textbf{30},
3356-3358 (2005).

\bibitem{M15} Zhang S., Fan W., Panoiu N. C., Malloy K. J., Osgood R. M.,
Brueck S. R. J. \emph{Demonstration of Near-Infrared Negative-Index
Materials.} arXiv. physics/0504208

\bibitem{M16} Podolskiy V.A., Narimanov E.E. \emph{Non-magnetic left handed
material}. arXiv:physics/0405077.

%
%

\bibitem{M23} Pendry J.B. \textit{Contemporary Physics} \textbf{45}, 191-202
(2004).

\bibitem{M24} Ramakrishna S A. \textit{Rep.Prog.Phys.} \textbf{68}, 449-521
(2005).

\bibitem{R1} Shen, Y.R.: \emph{The principles of non-linear optics}, John
Wiley \& Sons, New York, Chicester, Brisbane, Toronto, Singapore, 1984.

\bibitem{R2} Yariv, A., and Yeh, P.: \emph{Optical waves in crystals}. John
Wiley \& Sons, New York, Chicester, Brisbane, Toronto, Singapore, 1984.

\bibitem{R3} Malomed B., \emph{Solitons in optical media with quadratic
nonlinearity}, Nonlinear Science at the Dawn of the 21st Cen-tury (ed. by
P.L. Christiansen, M.P. Sorensen, and A.C. Scott), pp. 247-262, Springer,
2000.

\bibitem{R4} Buryak A.V. , Di Trapani P., Skryabin D.V., Trillo S., \emph{%
Optical solitons due to quadratic nonlinearities: from basic physics to
futuristic applications}, \textit{Physics Reports} \textbf{370}, 63-235
(2002).

\bibitem{R25} Agranovich V.M., Shen Y.R., Baughman R.H., Zakhidov A.A., 
\textit{Phys.Rev. }\textbf{B69}, 165112 (2004).

\bibitem{R26} Agranovich V.M., Shen Y.R., Baughman R.H., Zakhidov A.A. 
\textit{J. Luminescence} \textbf{110}, 167 (2004).

\bibitem{R5} Shadrivov I.V., Zharov A.A., Kivshar Yu.S.
arXiv:physics/0506092 v1 10 (2005); \textit{J.Opt.Soc.Amer.} \textbf{B23},
529-534 (2006).

\bibitem{R6} Popov A.K., Shalaev V.M., \emph{Negative-Index Metamaterials:
Second-Harmonic Generation, Manley-Rowe Relations and Parametric
Amplification} . arXiv: physics/0601055 v1 10 Jan 2006

\bibitem{R30} Kivshar Yu.S. and G.P. Agrawal. \emph{Optical Solitons: From
Fibers to Photonic Crystals}, (Academic, San Diego, 2003).

\bibitem{R33} He H., Werner M.J., Drummond P.D., \textit{Phys.Rev.} \textbf{%
E54}, 896 (1996).

\bibitem{R34} Kartashov Ya.V., Vysloukh V.A., Torner L. \textit{Phys.Rev.} 
\textbf{E67}, 066612 (2003).

\bibitem{R35} Kartashov Ya.V., Egorov A.A., Zelenina A.S., Vysloukh V.A.,
Torner L. \textit{Phys.Rev.} \textbf{E68}, 046609 (2003).

\bibitem{R36} D'Aguanno G., Mattiucci N., Scalora M., Bloemer M.J. \textit{%
Phys.Rev.Lett.} \textbf{93}, 213902 (2004).

\bibitem{R37} Shadrivov I.V., Kivshar Yu.S. \textit{J.Opt. A: Pure Appl.
Opt. }\textbf{7}, S68 (2005).

\bibitem{R40} Karamsin Yu.N., Sukhorukov A.P., \textit{Zh.Exp.Teor.Fis.} 
\textbf{68}, 834 (1975).

\bibitem{R41} Hayata K., Koshiba M. \textit{Phys.Rev.} \textbf{A50}, 675
(1994)

\bibitem{R42} Werner M.J., Drummond P.D. \textit{Opt. Lett.} \textbf{19},
613 (1994).
\end{thebibliography}
\end{document}